\documentstyle [trans] {rspublic}

\input{psfig}

\def\GeV{\,{\rm GeV}}

\def\sec{\,{\rm sec}}
\def\Gyr{\,{\rm Gyr}}

\def\rcm{\,{\rm cm}}

\def\Mpc{\,{\rm Mpc}}

\def\eV{{\,\rm eV}}

\def\cmm2{{\,\rm cm^{-2}}}
\def\cm2{{\,{\rm cm}^2}}
\def\cmm3{{\,{\rm cm}^{-3}}}
\def\gcmm3{{\,{\rm g\,cm^{-3}}}}
\def\kms{\,{\rm km\,s^{-1}}}

\def\mpl{{m_{\rm Pl}}}

\def\la{\mathrel{\mathpalette\fun <}}
\def\ga{\mathrel{\mathpalette\fun >}}
\def\fun#1#2{\lower3.6pt\vbox{\baselineskip0pt\lineskip.9pt
  \ialign{$\mathsurround=0pt#1\hfil##\hfil$\crcr#2\crcr\sim\crcr}}}

\begin {document}

\title [Structure from Quantum Fluctuations] {Large-scale Structure
from Quantum Fluctuations in the Early Universe}
\author [M. S. Turner] {Michael S. Turner}
\affiliation {Departments of Astronomy \& Astrophysics and of Physics\\
Enrico Fermi Institute \\
The University of Chicago \\ 5640 South Ellis Avenue \\
Chicago IL~~60637-1433 USA\\ and \\
Theoretical Astrophysics \\
Fermi National Accelerator Laboratory \\
Batavia IL~~60510-0500 USA}
\maketitle

\maketitle
\label {firstpage}
\begin {abstract}

A better understanding of the formation of large-scale structure
in the Universe is arguably the most pressing question in cosmology.
The most compelling and promising theoretical paradigm, Inflation
+ Cold Dark Matter, holds that the density inhomogeneities
that seeded the formation of structure in the Universe
originated from quantum fluctuations
arising during inflation and that the bulk of the dark matter exists as
slowing moving elementary particles (`cold dark matter') left over
from the earliest, fiery moments.  Large redshift surveys (such as
the SDSS and 2dF) and high-resolution measurements of
CBR anisotropy (to be made
by the MAP and Planck Surveyor satellites) have the potential to
decisively test Inflation + Cold Dark Matter and to open
a window to the very early Universe and fundamental physics.  

\end {abstract}

\section{From Quark Soup to Large-scale Structure}

The hot big-bang cosmology is so successful that for two decades
it has been called {\it the} standard cosmology (see e.g.,
Peebles 1993 or Kolb \& Turner 1990).  It provides an accounting
of the Universe from a fraction of a second after the beginning
when the Universe was a hot, smooth soup of quarks and leptons to the
present, some $13\Gyr$ later.  The observational foundation of
the standard cosmology rests upon
three strong pillars:  the expansion of the Universe; the cosmic
microwave background radiation (CBR); and the abundance pattern
of the light elements, D, $^3$He, $^4$He, and $^7$Li,
produced seconds after the bang (see e.g., Peebles {\it et al.}, 1991).

\begin{figure}
\centerline{\psfig{figure=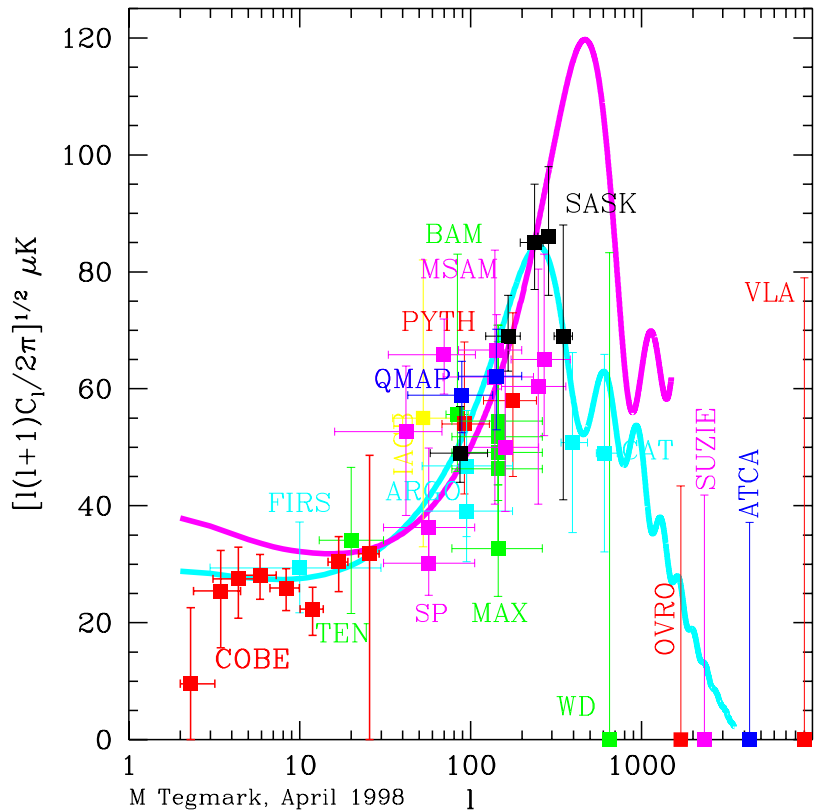,width=4.2in}}
\caption{Summary of current CBR anisotropy measurements, where
the temperature variation across the sky has been expanded
in spherical harmonics, $\delta T(\theta , \phi ) = \sum_i a_{lm}Y_{lm}$
and $C_l \equiv \langle |a_{lm}|^2\rangle$.  The curves
illustrate CDM models with $\Omega_0 = 1$ and $\Omega_0 =0.3$.
Note the preference of the data for a flat Universe
(Figure courtesy of M. Tegmark).}
\label{fig:cbr_today}
\end{figure}

In contrast to the early Universe, the Universe today abounds with structure:
galaxies, clusters of galaxies, superclusters, voids and great walls
of galaxies stretching across the sky.  According to the standard
cosmology, all this structure evolved by gravitational amplification
of small density inhomogeneities over the past $13\,$Gyr or so.
The detection of 30 microKelvin variations in the temperature of
the CBR between points on the sky separated by
$10^\circ$ by the DMR
instrument on NASA's COBE satellite (Smoot {\it et al.}, 1992)
gave the first evidence for the
existence of these density perturbations, and further, showed they
were of the size needed to account for the observed structure.
The COBE results have been followed up many other independent
detections on angular scales from $20^\circ$ down to a fraction of a degree
(Bennett {\it et al.}, 1997), summarized in Fig.~\ref{fig:cbr_today}.

While the standard cosmology leaves a number of fundamental questions
unexplained -- the matter/antimatter asymmetry, origin of the smoothness
and flatness of the Universe, nature of the big bang itself -- the
most pressing question involves the initial data for structure formation:
the nature and origin of the density inhomogeneities and the quantity
and composition of matter and energy in the Universe.  Because of powerful and
expansive theoretical ideas and an impending avalanche of data,
cosmology is poised for a major advance on this front,
and with it fundamental physics,
because the most promising ideas are inspired by speculations
about elementary particle physics at very high energies and short distances.

\subsection{Origin of inhomogeneity}

The two most promising ideas for the origin of the seed inhomogeneities
are cosmological inflation and cosmological symmetry breaking
phase transitions.  While both ideas involve the physics of
the early Universe, they are orthogonal, conceptually
and technically.  According to inflation, quantum-mechanical fluctuations
in the scalar field driving inflation lead to density perturbations,
which then become fluctuations in the
local curvature of the Universe during the inflationary epoch.
The other early-Universe alternative involves the production of
topological defects in a symmetry-breaking phase transition around
$10^{-36}\,$sec after the beginning.  The defects themselves  --
monopoles, cosmic strings, or textures -- do not directly lead
to density perturbations on astrophysically interesting scales.
Rather, the conversion of energy into defects causes a
pressure perturbation which propagates outward, and
much later on, leads to a perturbation in the matter density.  (This
type of density perturbation is called isocurvature.)  I will focus
exclusively on the inflationary scenario; the defect scenario is
discussed by Turok (1998).

\subsection{Nature of the matter and energy in the Universe}

The other crucial issue is the quantity and composition of matter and
energy in the Universe.  The amount of matter that clusters can be measured
in a variety of ways:  galaxy and cluster mass-to-light ratios,
peculiar motions of galaxies, the cluster baryon fraction, and the
shape of the power spectrum of density perturbations.  At present
all methods are consistent with $\Omega_M \simeq 0.35\pm 0.07$ (Dekel
{\it et al.}, 1997; Bahcall {\it et al.}, 1993; Willick {\it et al.}, 1997);
however, the remaining systematic uncertainties are such
that it is probably not possible to rule out $\Omega_M$ as small as 0.1
or as large as 1.  Matter in the form of stars and closely related
material contributes a tiny fraction of this,
$\Omega_{\rm lum} \simeq 0.003h^{-1}$.
The fact that $\Omega_{\rm lum} \ll \Omega_M$ implies that
most of the matter in the Universe is dark and is only revealed
by its gravitational effects.  ($\Omega_i$ is the fraction
of critical density contributed by component $i$, and
$h=H_0/100\kms \Mpc^{-1}$.  Current measurements of the Hubble
constant imply $h=0.65 \pm 0.07$).

The abundances of the light elements produced seconds after the
bang depends upon the density of ordinary matter (baryons); using
the recently determined ratio of deuterium-to-hydrogen in high-redshift
hydrogen clouds (Burles \& Tytler, 1998a and 1998b),
the theory of BBN implies that $\Omega_B = (0.02\pm 0.002)
h^{-2}\simeq 0.05$.  (This lies within the larger concordance
interval previously determined from the abundances of all the
light elements; see Schramm \& Turner, 1998.)

Since $\Omega_B \gg \Omega_{\rm lum}$, big-bang nucleosynthesis
implies that most of the baryons in the Universe are `dark' --
that is, not in the form of bright stars and closely related material
(plausible possibilities include diffuse hot and/or warm gas,
or dark stars).  Further, the fact that $\Omega_B$ is
significantly smaller than $\Omega_M$
strongly indicates that most of the matter is something other than baryons.
Elementary-particle physics provides three plausible particle
candidates:  light neutrinos; an axion of mass around $10^{-5}\,$eV,
and a neutralino of mass between 10\,GeV and 500\,GeV (see e.g.,
Turner 1993b; Jungman {\it et al.}, 1996).  The axion
and the neutralino have a predicted abundance today that is comparable
to the critical density; for neutrinos, whose number density today
is $113\cmm3$, a mass of order $30\eV$ corresponds to the critical
density.  All three possibilities are predictions made by
theories that attempt to go beyond the standard model of
particles and unify the forces and particles of Nature.

The total energy density is less well known.
Expressed as a fraction of the
critical density and denoted by $\Omega_0 = \sum_i \Omega_i$
($i=$ baryons, particle dark matter, vacuum energy, ....), it
is related to the spatial curvature,
\begin{equation}
R_{\rm curv}^2 = H_0^{-2}/|\Omega_0 -1| .
\end{equation}
The amount of dark matter implies that $\Omega_0$ must be greater than 0.2,
and the age of the Universe and the anisotropy
of the CBR constrain $\Omega_0$ to be not much greater than 1.
The most powerful measure of the curvature is the position
of the first acoustic (or Doppler) peak in the angular
power spectrum of CBR anisotropy:
$l_{\rm Doppler} \sim 220/\sqrt{\Omega_0}$.\footnote{The position
of the first acoustic (Doppler) peak also depends upon the
composition of the matter and energy density, e.g., the
presence or absence of a cosmological constant.  This dependence
is much less important.  If the density perturbations are
isocurvature, the Doppler peak is shifted to larger $l$.}
Current measurements are consistent with $\Omega_0=1$; see
Fig.~\ref{fig:cbr_today} and Hancock {\it et al.} (1997).
Ongoing measurements of anisotropy around
$l\sim 200$ (angular scale $\theta \sim 100^\circ / l
\sim 0.5^\circ$) may soon settle the question.

If $\Omega_0 =1$ and $\Omega_M = 0.3$ there is a
third dark-matter puzzle:  What is the nature of the component
of energy that does not clump with matter and is nearly
uniformly distributed?  To avoid clumping the `X-component'
($\Omega_X = 1 -\Omega_M \sim 0.7$)
must be relativistic (Turner \& White, 1997); however, relativistic
particles per se are out, because they lead to CBR anisotropy that
is inconsistent with current data (Lopez {\it et al.}, 1998)
and a Universe that is too youthful (for a radiation-dominated
Universe $t_0 = {1\over 2}H_0^{-1}$, rather than ${2\over 3}H_0^{-1}$
which pertains for a matter-dominated Universe).

The remaining possibility is that the smooth
component has negative pressure (is elastic) that is comparable
in magnitude to its energy density, $p_X \la
-\rho_X/3$.  Plausible examples include:  a cosmological
constant (or vacuum energy) with $p_X = -\rho_X$ (Turner {\it et al.}, 1984;
Peebles, 1984; Efstathiou {\it et al.}, 1990),
a network of light, frustrated defects
(e.g., strings in which case $p_X
=-\rho_X/3$; Vilenkin, 1984; Spergel \& Pen, 1997),
and an evolving scalar field (called quintessence
by some) with a time-varying relation between pressure and energy
density, $\rho = {1\over 2}\dot\phi^2 + V(\phi )$
and $p = {1\over 2}\dot\phi^2 - V(\phi )$ (Freese {\it et al.}, 1987;
Ozer \& Taha, 1987; Ratra \& Peebles, 1988; Bloomfield-Torres
\& Waga, 1996; Coble {\it et al.}, 1996; Caldwell {\it et al.}, 1998).

A smooth component does not
reveal its presence in dynamical measurements and is
difficult to detect.  It does have a striking signature:
an accelerated (rather than decelerated) expansion rate,
reflected in Sandage's deceleration parameter,
\begin{equation}
q_0 \equiv -{(\ddot R / R)\over H_0^2} = {\Omega_M\over 2}
        + {\Omega_X \over 2}\,[1 + 3p_X/\rho_X] < 0
\end{equation}
where $R(t)$ is the cosmic-scale factor.
Recent measurements of the magnitude -- redshift relation
for supernovae of type Ia (SNe1a) indicate
accelerated expansion ($q_0<0$), with $\Omega_X \sim
0.6$ and $p_X/\rho_X \sim -1$ (Riess {\it et al.}, 1998; Perlmutter
{\it et al.}, 1997).

In ending this brief review of the quantity and composition
of matter and energy in the Universe, I cannot resist commenting that, for
the very first time, we have a prima facie case for a complete
and consistent accounting:  The Doppler peak is telling us
that $\Omega_0 =1$; dynamical measurements indicate $\Omega_M
\sim 1/3$; and SNe1a indicate that $\Omega_X \sim 2/3$.  And
further, the picture that has emerged is consistent with inflation,
our most promising scenario for extending the standard cosmology.
If this turns out to be correct, 1998 will be remembered as a
turning point in our understanding of the Universe.

\section{From Quantum Fluctuations to Large-scale Structure}

Inflation has revolutionized the way cosmologists view the Universe
and provides the current working hypothesis for extending the
standard cosmology.  It explains how a region of size much, much
greater than our Hubble volume could have become smooth and flat
without recourse to special initial conditions (Guth 1981), as well as
the origin of the density inhomogeneities
needed to seed structure (Hawking, 1982; Starobinsky, 1982;
Guth \& Pi, 1982; and Bardeen {\it et al.}, 1983).
Inflation is based upon
well defined, albeit speculative physics -- the semi-classical evolution
of a weakly coupled scalar field -- and this physics may well
be connected to the unification of the particles and forces of Nature.

On the negative side, while there are numerous working models
of inflation, motivated by a variety of
concerns -- supersymmetry, superstrings, grand unification and simplicity --
there is no standard model of inflation.
And a disquieting technical point, in all models of inflation the scalar
field that drives inflation must have a very flat potential and
must be very weakly coupled to other fields.  Most particle physicists
find this displeasing or, at the very least, begging for further explanation.
The extreme flatness and weak coupling trace directly to the requirement
of producing density perturbations of amplitude $10^{-5}$ (for
recent reviews of inflation see e.g., Turner, 1997a or Lyth \& Riotto, 1998).

It would be nice if there were a standard model of inflation, but
there isn't.   What is important, is that almost all inflationary
models make three very testable predictions:  flat
Universe,\footnote{It is possible,
by the introduction of additional scalar fields and fine tuning,
to evade the flatness prediction; this author
still considers flatness to be a robust prediction of inflation.
For another opinion,
see Bucher {\it et al.} (1995), Linde \& Mezhlumian (1995), or
Turok (1998).} nearly scale-invariant
spectrum of Gaussian density perturbations, and nearly scale-invariant
spectrum of gravitational waves.  These three predictions allow the
inflationary paradigm to be decisively tested.  While the gravitational
waves are an extremely important test, I do not have
space to mention them again here (see e.g., Turner 1997c).

The difference between different models of inflation lies in the
scalar-field potential; once the scalar-field potential
is specified, the story is the same.  Inflation begins with the
scalar field displaced from the minimum of its potential (for
whatever reason); as it evolves toward the potential-energy minimum
the scalar-field potential energy drives a nearly exponential
expansion.  In most models, the time required
to evolve to the minimum is many hundreds or thousands of
Hubble times, during which the scale factor of the Universe grows by an
enormous factor.  When the scalar field nears the minimum of its
potential, its evolution accelerates and it rapidly oscillates
about the minimum.  `The graceful exit' from the inflationary era
occurs as the original potential energy, which now resides
in coherent scalar-field oscillations, decays into relativistic
particles, which through interactions eventually thermalize,
creating the heat of the hot big-bang model.

The tremendous expansion that occurs during inflation
is key to its beneficial effects and robust predictions:
A small, subhorizon-sized bit
of the Universe can grow large enough to encompass the entire
observable Universe and much more.  The same small bit of
the Universe is smaller than its radius of
curvature and appears flat; this relationship is unaffected
by the expansion since then and so the Hubble radius today is
much, much smaller than the curvature radius, implying $\Omega_0
=1$ (recall, $R_{\rm curv} = H_0^{-1}/|\Omega_0 -1|^{1/2}$).
Lastly, the tremendous expansion stretches quantum fluctuations
on truly microscopic scales ($\la 10^{-23}\rcm$)
to astrophysical scales ($\ga \Mpc$).

The accelerated expansion associated
with inflation is crucial.  If, and
only if, the expansion accelerates (i.e., $\ddot R > 0$),
can a comoving scale begin
smaller than the horizon and grow larger.  In the standard cosmology,
the expansion is always decelerating, and all comoving scales (e.g., the
scale corresponding to the presently observable Universe)
begin larger than the horizon scale (set by the inverse of the
expansion rate $H^{-1}$) and then cross inside the horizon.  Thus,
objects from galaxies to the presently observable Universe were
much larger than the horizon during the earliest moments and
outside the sphere of causal influence.  Inflation
changes that:  these objects begin smaller than the horizon
where microphysics can affect them and then cross outside the
horizon during inflation.

Accelerated expansion makes it kinematically possible to create
density inhomogeneities on astrophysical interesting scales,
and the quantum fluctuations associated with the deSitter space of
accelerated expansion provide the dynamical mechanism.
Quantum fluctuations in the scalar field that drives inflation, whose
amplitude is set by the Gibbons--Hawking temperature $H/2\pi$,
lead to energy density fluctuations $\delta \rho = V^\prime \Delta \phi
=V^\prime H /2\pi$.
As each scale, from galaxies to clusters to the present Hubble scale,
crosses outside the horizon, these perturbations become fluctuations in the
curvature of the Universe.

The curvature perturbations created by inflation
are characterized by two important features: 1) they are
almost scale-invariant, which refers to the fluctuations in
the gravitational potential being independent
of scale -- and not the density perturbations themselves;
2) because they arise from fluctuations in an essentially noninteracting
quantum field, their statistical properties are that of
a Gaussian random field.

\begin{figure}
\centerline{\psfig{figure=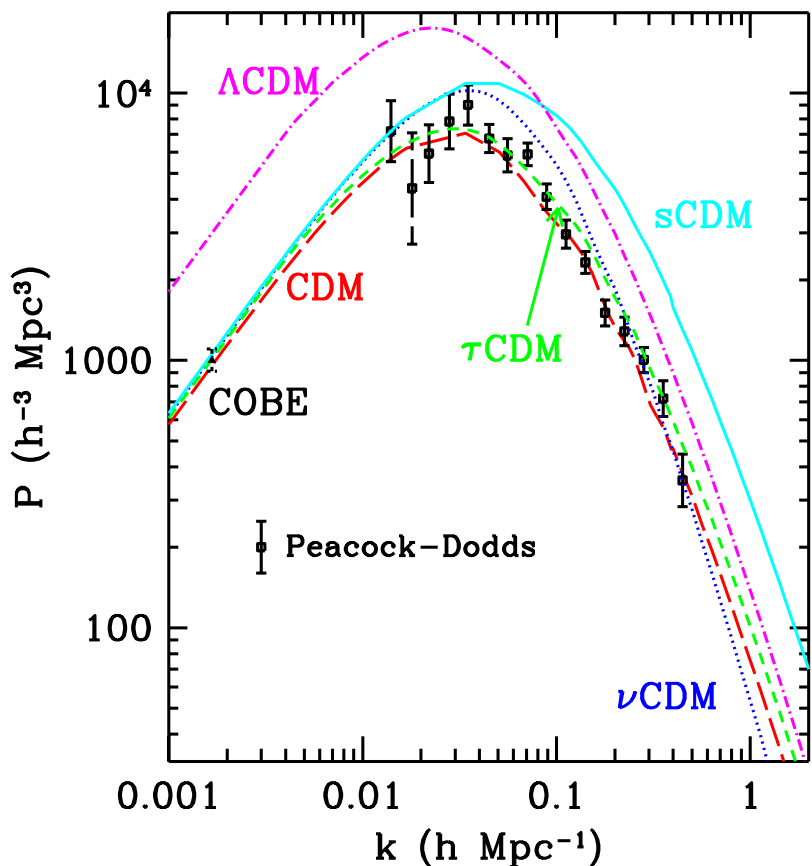,width=4.2in}}
\caption{Summary of measurements of the present power spectrum
derived from redshift surveys (Peacock \& Dodds, 1994) and
the predictions of different COBE-normalized CDM models.
Because of the possibility that light does not faithfully
trace mass (i.e., biasing), the shape of the spectrum is
most important in constraining models.}
\label{fig:ps_today}
\end{figure}

Scale invariance specifies the dependence of the spectrum of density
perturbations upon scale.  The normalization (overall amplitude) depends
upon the specific inflationary model (i.e., scalar-field potential).
Ignoring numerical factors for the moment, the fluctuation amplitude
is given by:  $\delta \phi\simeq (\delta \rho /\rho )_{\rm HOR} \sim
V^{3/2}/m_{\rm PL}^3 V^\prime$.
(The amplitude of the density perturbation on a given scale
at horizon crossing is equal to the fluctuation in the gravitational
potential $\delta \phi$.)  To be consistent with the COBE
measurement of CBR anisotropy
on the $10^\circ$ scale, $\delta \phi$ must be around $2\times 10^{-5}$.
Not only did COBE produce the first evidence for the existence of
the density perturbations that seeded all structure, but also,
for a theory like inflation that predicts
the shape of the spectrum of density perturbations,
it provides the overall normalization that
fixes the amplitude of density perturbations on all scales
(see Fig.~\ref{fig:ps_today}).
The COBE normalization began precision testing of inflation.

\section{Testing Inflation + CDM in the Era of Precision Cosmology}

The inflationary predictions of a flat Universe and scale-invariant
density perturbations, together with the failure of the
hot dark matter theory of structure formation,
make cold dark matter (CDM) a key prediction
and a powerful means of testing inflation.
The key elements of CDM are:  1) Gaussian scale-invariant density
perturbations; and 2) dark matter whose primary constituent
is slowly-moving, very weakly interacting particles such as axions
or neutralinos.  CDM is hierarchical in the sense that
structure forms from the `bottom up' -- galaxies (at redshifts of a few),
followed by clusters of galaxies (redshifts of one or less) and
finally superclusters (today) (see e.g.,
Blumenthal {\it et al.}, 1984).\footnote{If the bulk
of the dark matter is ``hot'' -- that is fast moving particles
such as 30\,eV neutrinos -- then structure forms from the `top down,'
with superclusters forming first and fragmenting into galaxies.
This is because hot dark matter particles can stream from regions
of high density to regions of low density and erase density perturbations
on scales smaller than superclusters.  Since the pioneering work
of White {\it et al.} (1983), hot dark matter has been disfavored
because galaxies form too late.  Since we now know that the bulk
of galaxies formed at redshifts of a few and superclusters are
only forming today (Steidel, 1998) the hot dark
matter scenario is completely incompatible with observations.}

CDM is generally consistent with the key tests that have been
carried out thus far:  anisotropy of the CBR on angular scales
from less than a degree to $100^\circ$, measurements of the
distribution of galaxies today, and studies of the evolution
of galaxies and clusters (see e.g., Steidel, 1998).  This is no
mean feat; at present, CDM is the {\em only} theory for structure
formation that is still viable:  the theories based
upon defects as the seeds for structure
are strongly disfavored by a combination
of CBR anisotropy and the power spectrum of inhomogeneity today
(Pen {\it et al.}, 1997; Allen {\it et al.} 1997) and Peebles' ``baryon only''
model (Peebles, 1987) with isocurvature perturbations (PIB)
was ruled out by CBR anisotropy several years ago.

As we look forward to the abundance (avalanche!) of high-quality observations
that will test Inflation + CDM, we have to make sure the predictions
of the theory match the precision of the data.  In
so doing, CDM + Inflation becomes a ten (or more) parameter
theory.  For astrophysicists, and especially cosmologists,
this is daunting, as it may seem that a ten-parameter
theory can be made to fit any set of observations.  This is
not the case when one has the quality and quantity of data
that will be coming.  The standard model of particle physics offers
an excellent example:  it is
a nineteen-parameter theory and because of the high-quality of
data from experiments at Fermilab's Tevatron, SLAC's SLC,
CERN's LEP and other facilities it has been rigorously tested
and the parameters measured to a precision of better than 1\%
in some cases.  My worry as an inflationist is not that many different
sets of parameters will fit the upcoming data, but rather that
no set of parameters will!

In fact, the ten parameters of CDM + Inflation
are an opportunity rather than a curse:  Because the parameters
depend upon the underlying inflationary model and fundamental
aspects of the Universe, we have the very real possibility of learning
much about the Universe and inflation.  The ten parameters
can be organized into two groups:  cosmological
and dark-matter (Dodelson {\it et al.}, 1996).

\smallskip
\centerline{\it Cosmological Parameters}
\vspace{3pt}
\begin{enumerate}

\item $h$, the Hubble constant in units of $100\kms\Mpc^{-1}$.

\item $\Omega_Bh^2$, the baryon density.  Primeval deuterium
measurements and together with the theory of BBN imply:
$\Omega_Bh^2 = 0.02 \pm 0.002$.

\item $n$, the power-law index of the scalar density perturbations.
CBR measurements indicate $n=1.1\pm 0.2$; $n=1$ corresponds to
scale-invariant density perturbations.  Several popular
inflationary models predict $n\simeq 0.95$; range of predictions
runs from $0.7$ to $1.2$ (Lyth \& Riotto, 1996; Huterer \& Turner, 1998).

\item $dn/d\ln k$, ``running'' of the scalar index with comoving scale
($k=$ wavenumber).  Inflationary models predict a value of
${\cal O}(\pm 10^{-3})$ or smaller (Kosowsky \& Turner, 1995).

\item $S$, the overall amplitude squared of density perturbations,
quantified by their contribution to the variance of the
CBR quadrupole anisotropy.

\item $T$, the overall amplitude squared of gravity waves,
quantified by their contribution to the variance of the
CBR quadrupole anisotropy.  Note, the COBE normalization determines
$T+S$ (see below).

\item $n_T$, the power-law index of the gravity wave spectrum.
Scale-invariance corresponds to $n_T=0$; for inflation, $n_T$
is given by $-{1\over 7}{T\over S}$.

\end{enumerate}

\smallskip
\centerline{\it Dark-matter Parameters}
\vspace{3pt}

\begin{enumerate}

\item $\Omega_\nu$, the fraction of critical density in neutrinos
($=\sum_i m_{\nu_i}/90h^2$).  While the hot dark matter theory of structure
formation is not viable, it is possible that a small fraction of
the matter density exists in the form of neutrinos.
Further, small -- but nonzero -- neutrino masses are
a generic prediction of theories that unify the
strong, weak and electromagnetic interactions.\footnote{As this
article went to press, the Super-Kamiokande Collaboration presented
evidence that the at least one of the neutrino species has a
mass of greater than about 0.1\,eV, based upon the deficit
of atmospheric muon neutrinos (Kajita, 1998).}

\item $\Omega_X$, the fraction of critical density in a smooth component
of unknown composition and negative pressure ($w_X \la -0.3$).  There is
mounting evidence for such a component, with the simplest example being
a cosmological constant ($w_X = -1$).

\item $g_*$, the quantity that counts the number of ultra-relativistic
degrees of freedom (around the time of matter-radiation
equality).  The standard cosmology/standard
model of particle physics predicts $g_* = 3.3626$ (photons in the
CBR + 3 massless neutrino species with temperature $(4/11)^{1/3}$
times that of the photons).  The amount of radiation controls when
the Universe became matter dominated and thus affects the present
spectrum of density inhomogeneity.

\end{enumerate}

Since $\Omega_0 =1.0$ is taken to be an inflationary
prediction, $\Omega_{\rm CDM} = 1 - \Omega_\nu - \Omega_B - \Omega_X$.
Additional parameters can be added (e.g., $\Omega_0$, $w_X$, the
epoch of reionization).   Bond's list totals
nineteen, coincidentally equal to the number of parameters
in the standard model of particle physics
(Bond \& Jaffe, 1998).  The main point is that testing inflation + CDM
requires precision predictions, which in turn, depend on ten or so parameters.

As mentioned, the parameters involving density and gravity-wave
perturbations depend directly upon the inflationary potential.
In particular, they can be expressed in terms of the potential
and its first three derivatives (see e.g., Turner, 1997a):

\begin{eqnarray}
S  & \equiv & {5\langle |a_{2m}|^2\rangle \over 4\pi} \simeq
         2.2\,{V_*/m_{\rm Pl}^4 \over (m_{\rm Pl} V_*^\prime /V_*)^2}\\
n -1 & = & -{1\over 8\pi}\left({\mpl V^\prime_* \over V_*} \right)^2
        + {\mpl \over 4\pi}\left( {\mpl V_*^{\prime}\over V_*} \right)^\prime \\
{dn\over d\ln k} & = &
-{1\over 32\pi^2}\left({{m_{\rm Pl}}^3V_*^{\prime\prime\prime}\over
        V_*}\right)
        \left({{m_{\rm Pl}} V_*^\prime\over V_*}\right) \nonumber\\
     & & \hspace*{1cm} + {1\over 8\pi^2}
        \left({{m_{\rm Pl}}^2V_*^{\prime\prime}\over V_*}\right)
        \left({{m_{\rm Pl}} V_*^\prime\over V_*}\right)^2
        - {3\over 32\pi^2}\left({m_{\rm Pl}} {V_*^\prime\over V_*}\right)^4 \\
T  & \equiv & {5\langle |a_{2m}|^2\rangle \over 4\pi} =
        0.61 (V_*/m_{\rm Pl}^4) \\
n_T & = & -{1\over 8\pi} \left( {\mpl V^\prime_* \over V_*} \right)^2
\label{inflateeqns}
\end{eqnarray}
where $V(\phi )$ is the inflationary potential, prime denotes $d/d\phi$,
and $V_*$ is the value of the scalar potential when the present horizon
scale crossed outside the horizon during inflation.
These expressions are
given to lowest-order in the deviation from scale invariance
(i.e., $n-1$ and $n_T$),
and assume a matter-dominated Universe today; the next-order
corrections have been calculated (Liddle \& Turner, 1994)
and the analogous expressions, including the possibility of a cosmological
constant, have been computed (Turner \& White, 1996).

Bunn \& White (1997) have used the
COBE four-year dataset to determine $S$ as a function of
$T/S$ and $n-1$; they find
\begin{equation}
{V_*/m_{\rm Pl}^4 \over (m_{\rm Pl} V_*^\prime /V_*)^2}
        \left( = {S\over 2.2} \right) =
        (1.7\pm 0.2)\times 10^{-11} \,{\exp [-2.02(n-1)] \over
        \sqrt{1+{2\over 3}{T\over S}}}
\end{equation}
{}From which it follows that
\begin{equation}
V_* < 6\times 10^{-11} \mpl^4,
\end{equation}
equivalently, $V_*^{1/4} < 3.4\times 10^{16}\GeV$.
This indicates that inflation must involve
energies much smaller than the Planck scale.
(To be more precise, inflation could have begun at a much higher
energy scale, but the portion of inflation relevant for us, i.e.,
the last 60 or so e-folds, occurred at an energy scale much smaller
than the Planck energy.)

This normalization can also be expressed in terms of
the horizon-crossing amplitude for the comoving scale $k=H_0$:
\begin{equation}
\delta_H(k=H_0) \equiv \left[ {k^{3/2}|\delta_k| \over
        \sqrt{2\pi^2}} \right]_{k=H_0} = 1.9\times 10^{-5} \,
        {\exp [-1.01(n-1)]\over \sqrt{1+{2\over 3}\,{T\over S}}} .
\end{equation}
That is, for $n=1$ and $T/S=0$, the COBE normalization implies that
the horizon-crossing amplitude of density perturbations is about
$2\times 10^{-5}$.

Finally, it should be noted that the `tensor tilt,'
deviation of $n_T$ from 0, and the `scalar
tilt,' deviation of $n-1$ from zero, are not in general equal;
they differ by the rate of change of the steepness.
The tensor tilt and the ratio $T/S$ are related:
$n_T = -{1\over 7}{T\over S}$, which provides a potential
consistency test of inflation.

\subsection{Present status of Inflation + CDM}

\begin{figure}
\centerline{\psfig{figure=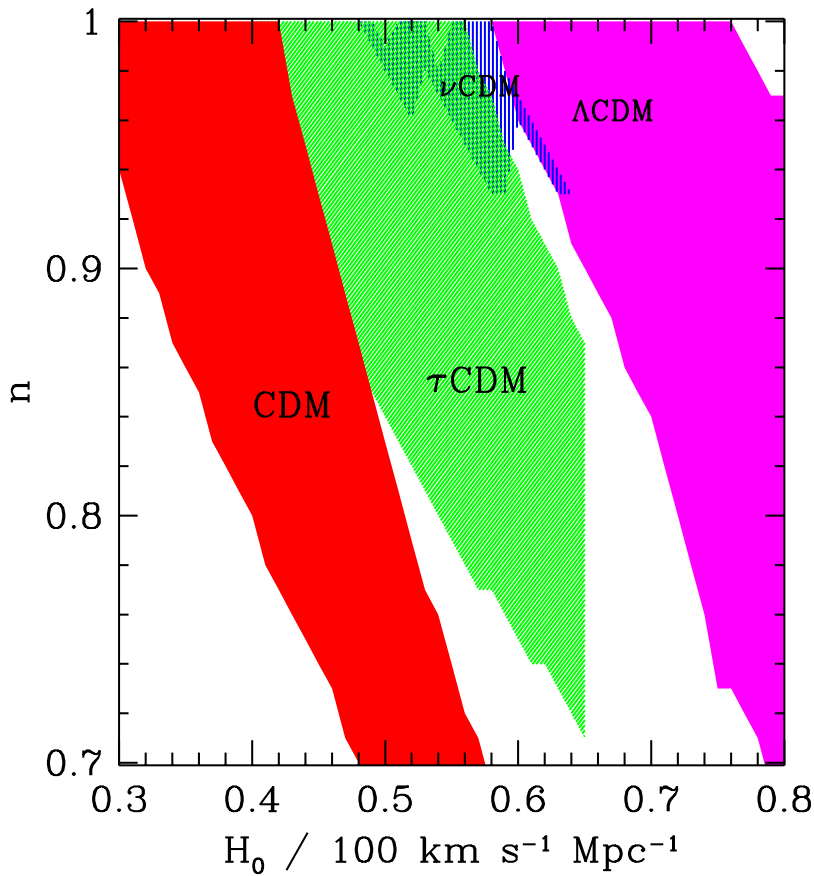,width=4.2in}}
\caption{Summary of viable CDM models, based upon
CBR anisotropy and determinations of the present
power spectrum of inhomogeneity (Dodelson {\it et al.}, 1996).}
\label{fig:cdm_sum}
\end{figure}

A useful way to organize the different CDM models is by their
dark-matter content; within each CDM family, the cosmological
parameters vary.  One list of models is:

\begin{enumerate}

\item sCDM (for simple):  Only CDM and baryons; no additional
radiation ($g_*=3.36$).  The original standard CDM is a member
of this family ($h=0.50$, $n=1.00$, $\Omega_B=0.05$), but is
now ruled out (see Fig.~\ref{fig:cdm_sum}).

\item $\tau$CDM:  This model has
extra radiation, e.g., produced by the decay of an unstable
massive tau neutrino (hence the name); here we take $g_* = 7.45$.

\item $\nu$CDM (for neutrinos):  This model has a dash of hot
dark matter; here we take $\Omega_\nu = 0.2$ (about 5\,eV
worth of neutrinos).

\item $\Lambda$CDM (for cosmological constant):  This model has
a smooth component in the form of a cosmological constant; here
we take $\Omega_\Lambda = 0.6$.

\end{enumerate}

Figure \ref{fig:cdm_sum} summarizes the viability of these different CDM models,
based upon CBR measurements and current determinations of
the present power spectrum of inhomogeneity (derived from
redshift surveys).   sCDM is only viable for low values of the
Hubble constant (less than $55\kms\Mpc^{-1}$) and/or
significant tilt (deviation from scale invariance); the region
of viability for $\tau$CDM is similar to sCDM, but shifted
to larger values of the Hubble constant (as large as
$65\kms\Mpc^{-1}$).  $\nu$CDM has an island of viability
around $H_0\sim 60\kms\Mpc^{-1}$ and $n\sim 0.95$.  $\Lambda$CDM
can tolerate the largest values of the Hubble constant.

\begin{figure}
\centerline{\psfig{figure=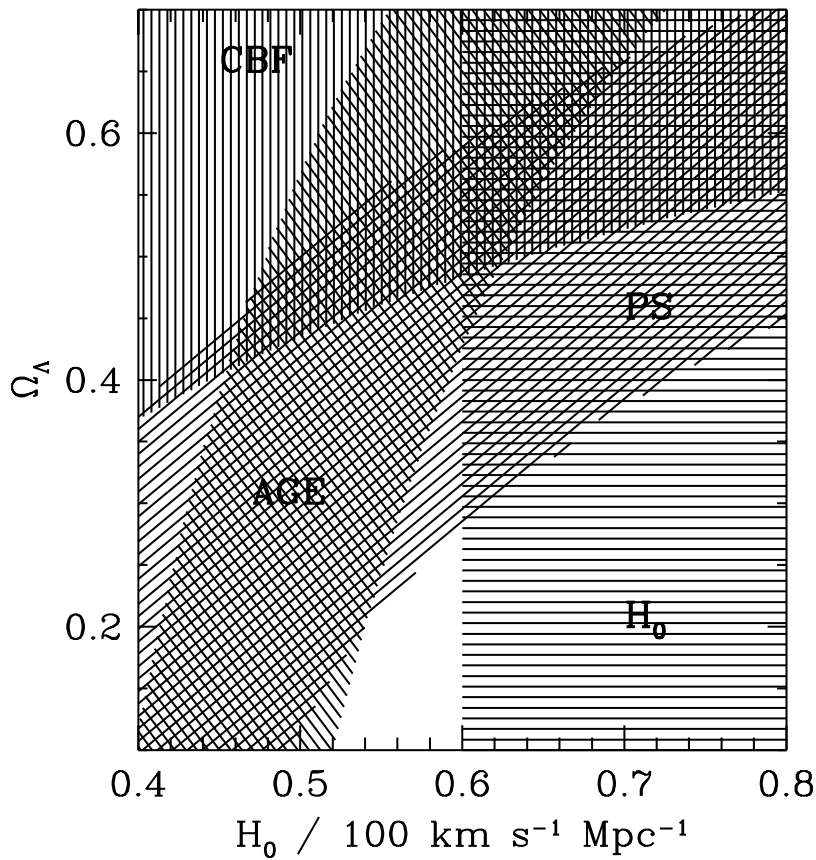,width=4.2in}}
\caption{Constraints used to determine the best-fit CDM model:
PS = large-scale structure + CBR anisotropy; AGE = age of the
Universe; CBF = cluster-baryon fraction; and $H_0$= Hubble
constant measurements.  The best-fit model, indicated by
the darkest region, has $h\simeq 0.60-0.65$ and $\Omega_\Lambda
\simeq 0.55 - 0.65$ (Krauss \& Turner 1995; Turner, 1997c).}
\label{fig:best_fit}
\end{figure}

Considering other relevant data too -- e.g.,
age of the Universe, determinations of $\Omega_M$,
measurements of the Hubble constant, and limits to
$\Omega_\Lambda$ -- $\Lambda$CDM emerges as the hands-down-winner
of `best-fit CDM model' (Krauss \& Turner, 1995;
Ostriker \& Steinhardt, 1995; Liddle {\it et al.}, 1996;
Turner, 1997b).  Moreover, not only
is it consistent with all the data (see Fig.~\ref{fig:best_fit}),
but also its `smoking gun signature,'
negative $q_0$, has apparently been confirmed (Riess {\it et al.}, 1998;
Perlmutter {\it et al.}, 1997).  Given the possible systematic uncertainties
in the SNe1a data and other measurements, it is premature to conclude that
$\Lambda$CDM is anything but the model to take aim at.

\section{Testing Inflation with Maps of the Universe}

Over the next decade two maps of the Universe with unprecedented
precision will be made.  The first, derived from high-resolution
(around $0.1^\circ$) measurements of the CBR
by NASA's MAP and ESA's Planck satellites, will provide a snapshot
of the Universe at a simpler time, 300,000\,yrs after the
beginning when the average level of inhomogeneity was much less than
1\%.  The second, derived from the more than one million galaxy
redshifts to be gathered by the SDSS and 2dF teams, will provide
an accurate picture of the structure that exists in the Universe today.
The two maps are complementary and together have great leverage
to settle the question of how structure in the Universe originated
as well as to probe cosmology and fundamental physics.

The SDSS and 2dF redshift surveys will probe the Universe on scales from
less than $1h^{-1}\Mpc$ to $500h^{-1}\Mpc$.
The structure that exists today depends
not only upon the primordial spectrum of inhomogeneity, but also
upon the composition of the dark matter, cosmological parameters
and the complicated astrophysical relationship between the present
distribution of light and mass.  CBR anisotropy probes the primeval
spectrum of inhomogeneity on scales from $10h^{-1}\Mpc$ to $10^4h^{-1}
\Mpc$.  Together, they will probe inhomogeneity over almost six
orders-of-magnitude in length.

The power of these two maps when used together
has been stressed by a number of authors
(see e.g., Eisenstein {\it et al.}, 1998).
I mention but a few examples.  CBR anisotropy
should determine $\Omega_0$ and $h$ to a precision of better than
1\%; large-scale structure can accurately determine $\Omega_M h$ (the
shape parameter).  Together, they determine accurately
$\Omega_M$, $\Omega_X$
and $h$.  The effect of a neutrino mass as small as a few tenths
of an eV should be detectable by a combination of redshift data and
CBR anisotropy (Hu {\it et al.}, 1998).  The two maps
will both probe inhomogeneity
on scales of $10h^{-1}\Mpc$ to around $500h^{-1}\Mpc$, which will
allow the mismatch between the distribution of light and mass
(biasing) to be addressed.  

\subsection{Looking `out' to see `in'}

Inflation and cold dark matter are a bold attempt to extend
our knowledge of the Universe to within $10^{-32}\sec$
of the bang.  The scenario is deeply rooted in fundamental
physics.  I am confident that redshift surveys, CBR anisotropy
and a host of other cosmological observations and laboratory
experiments will decisively test inflation + CDM.  Further,
I believe prospects for discriminating among the
different CDM models and models of inflation are excellent.
If inflation + CDM is shown to be correct, an important aspect of
the standard cosmology -- the origin and evolution of structure --
will have been resolved and a window to the early moments of the Universe
and physics at very high energies will have been opened.

While the window has not been opened yet, I would like to
end with one example of what one could
hope to learn.  As discussed earlier, $S$, $n-1$, $T/S$ and
$n_T$ are related to the inflationary potential and its first
two derivatives.  If one can measure the power-law
index of the density perturbations and the amplitudes of the density
and gravity-wave perturbations, one can recover the value of the potential
and its first two derivatives (see e.g.,
Turner 1993a; Lidsey {\it et al.} 1997)
\begin{eqnarray}
V_* & = & 1.65 T\, {m_{\rm Pl}}^4  , \\
V_*^\prime & = & \pm \sqrt{{8\pi \over 7}\,{T\over S}}\, V_*/{m_{\rm Pl}} , \\
V_*^{\prime\prime} & = & 4\pi \left[ (n-1) + {3\over 7} {T\over S} \right]\,
V_* /{m_{\rm Pl}}^2 ,
\end{eqnarray}
where the sign of $V^\prime$ is indeterminate (under the
redefinition $\phi \leftrightarrow
-\phi$ the sign changes).  If, in addition, the gravity-wave spectral index
can also be measured the consistency relation, $T/S = -7 n_T$,
can be used to test inflation.
Reconstruction of the inflationary scalar potential would
shed light both on inflation as well as physics at energies of the
order of $10^{14}\GeV$.  Already,
the success of Inflation + CDM is evidence for physics
beyond the standard model of particle physics.

\begin {thebibliography}{}

\item Allen, B., Caldwell, R. R., Dodelson, S.,
Knox, L., Shellard, E. P. S., \& Stebbins, A. 1997
Cosmic Microwave Background anisotropy induced by cosmic strings
on angular scales greater than about $15^\prime$.
{\it Phys. Rev. Lett.} {\bf 79}, 2624-2627.

\item Bahcall, N.A. {\it et al.} 1993 {\it Astrophys. J.} {\bf 415}, L17.

\item Bardeen, J.,  Steinhardt, P.J., \& Turner, M.S. 1983 Spontaneous creation of almost scale-free density perturbations in an inflationary universe.  
 {\it Phys. Rev. D} {\bf 28}, 679-693.

\item Bennett, C. {\it et al.} 1997 {\it Physics Today} Nov., 32.

\item Bloomfield-Torres, L.F. \& Waga, I. 1996 {\it Mon. Not. R. Astron. Soc.} {\bf 279}, 712.

\item Blumenthal, G. R., Faber, S. M., Primack, J. R., \& Rees, M. J. 1984 Formation of galaxies and large-scale structure with cold dark matter.  {\it Nature} {\bf 311}, 517.

\item Bond, J.R. \& Jaffe, A. 1998 in this volume.

\item Bucher, M., Goldhaber, A. S., \& Turok, N. 1995 Open universe from inflation. {\it Phys. Rev. D} {\bf 52}, 3314.

\item Bunn, E. F. \& White, M. 1997 The 4-year COBE normalization and large-scale structure. {\it Astrophys. J.} {\bf 480}, 6-21.

\item Burles, S. \& Tytler, D. 1998a The deuterium abundance toward Q1937-1009. {\it Astrophys. J.} {\bf 499}, 699-712.

\item Burles, S. \& Tytler, D. 1998b {\it Astrophys. J.} in press.

\item Caldwell, R., Dave, R., \& Steinhardt, P.J. 1998 Cosmological imprint of an energy component with general equation of state.  {\it Phys. Rev. Lett.} {\bf 80}, 1582-1585.

\item Coble, K., Dodelson, S., \& Frieman, J. A. 1996 Dynamical $\Lambda$ models of structure formation. {\it Phys. Rev. D} {\bf 55}, 1851.

\item Dekel, A. {\it et al.} 1997.  In {\it Critical dialogues in cosmology} (ed. N. Turok), p.175. Singapore:World Scientific.

\item Dodelson, S. {\it et al.} 1996 {\it Science} {\bf 274}, 69.

\item Efstathiou, G., Sutherland, W.J., \& Maddox, S. J. 1990 The cosmological constant and cold dark matter. {\it Nature} {\bf 348}, 705-707.

\item Eisenstein, D. J., Hu, W., \& Tegmark, M. 1998 Cosmic complementarity: H$_0$ and $\rm \Omega_m$ from combining CMB experiments and redshift surveys. {\it Astrophys. J.}, submitted (astro-ph/9805239).

\item Freese, K. {\it et al.} 1987 {\it Nucl. Phys. B} {\bf 287}, 797.

\item Guth, A.H. 1981 Inflationary universe: A possible solution to the horizon and flatness problems. {\it Phys. Rev. D} {\bf 23}, 347.

\item Guth, A.H. \& Pi, S.-Y. 1982 Fluctuations in the new inflationary universe. {\it Phys. Rev. Lett.}
{\bf 49}, 1110-1113.

\item Hancock {\it et al.} 1998 {\it Mon. Not. R. Astron. Soc.}, in press (astro-ph/9708254).

\item Hawking, S.W. 1982 The development of inequalities in a single bubble inflationary universe.  {\it Phys. Lett. B} {\bf 115}, 295-297.

\item Hu, W., Eisenstein, D. J., \& Tegmark, M. 1998 Weighing neutrinos with galaxy surveys. {\it Phys. Rev. Lett.} {\bf 80}, 5255-5258.

\item Huterer, D. \& Turner, M.S. 1998, in preparation.

\item Jungman, G. {\it et al.} 1996 {\it Phys. Rep.} {\bf 267}, 195.

\item Kajita, T. (for the Super-Kamiokande Collaboration) 1998,
presented at {\it Neutrino 98} (Takayama, Japan, June 4-9).

\item Kolb, E.W. \& Turner, M.S. 1990 {\it The early universe}. Redwood
City, CA:Addison-Wesley.

\item Kosowsky, A. \& Turner, M.S. 1995 CBR anisotropy and the running of the scalar spectral index.  {\it Phys. Rev. D} {\bf 52}, R1739.

\item Krauss, L. \& Turner, M.S. 1995 The cosmological constant is back.  {\it Gen. Rel. Grav.} {\bf 27}, 1137.

\item Liddle, A.R.  {\it et al.} 1996 {\it Mon. Not. R. astron. Soc.} {\bf 282}, 281.

\item Liddle, A.R. \& Turner, M.S. 1994 Second-order reconstruction of the inflationary potential. {\it Physical Review D} {\bf 50}, 758.

\item Lidsey, J. {\it et al.} 1997 {\it Rev. Mod. Phys.} {\bf 69}, 373.

\item Linde, A.D. \& Mezhlumian, A. 1995 Inflation with $\Omega \not= $ 1. {\it Phys. Rev. D} {\bf 52}, 6789-6804.

\item Lopez, R. E., Dodelson, S., Scherrer,
R. J., \& Turner, M. S. 1998 Probing unstable massive
neutrinos with current CMB observations. astro-ph/9806116.

\item Lyth, D. H. \& Riotto, A.
1998 Particle physics models of inflation and the cosmological
density perturbation. {\it Phys. Rep.}, in press (hep-th/9609431 and
9807278).

\item Ostriker, J.P. \& Steinhardt, P.J. 1995 {\it Nature} {\bf 377}, 600.

\item Ozer, M. \& Taha, M.O. 1987 {\it Nucl. Phys. B} {\bf 287}, 776.

\item Peacock, J.  \& Dodds, S.  1994 Reconstructing the linear power spectrum of cosmological mass fluctuations.  {\it Mon. Not. R. Astron. Soc.} {\bf 267}, 1020-1034.

\item Peebles, P.J.E. 1984 {\it Astrophys. J.} {\bf 284}, 439.

\item Peebles, P.J.E. 1987 Origin of the large-scale galaxy peculiar velocity field: A miniature isocurvature model.  {\it Nature} {\bf 327}, 210-211.

\item Peebles, P.J.E. 1993 {\it Principles of physical cosmology}. Princeton NJ:Princeton Univ. Press.

\item Peebles, P.J.E., Schramm, D. N., Turner, E. L., \& Kron, R.G. 1991 The case for the relativistic hot big bang cosmology. {\it Nature} {\bf 352}, 769-776.

\item Pen, U.-L. {\it et al.} 1997 {\it Phys. Rev. Lett.} {\bf 79}, 1611.

\item Perlmutter, S. {\it et al.} 1997 {\it B.A.A.S.} {\bf 5}, 1351.

\item Ratra, B. \& Peebles, P.J.E. 1988 Cosmological consequences of a rolling homogeneous scalar field. {\it Phys. Rev. D} {\bf 37}, 3406-3427.

\item Riess, A. {\it et al.} 1998 {\it Astron. J.}, in press.

\item Schramm, D.N. \& Turner, M.S. 1998 Big bang nucleosynthesis enters the precision era.  {\it Rev. Mod. Phys.} {\bf 70}, 303.

\item Smoot, G., {\it et al.} 1992 Structure in the COBE differential microwave radiometer first year maps. {\it Astrophys. J.} {\bf 396}, L1-L6.

\item Spergel, D. N. \& Pen, U.-L. 1997 String-dominated universe cosmology. {\it Astrophys. J.} {\bf 491}, L67.

\item Starobinsky, A. A. 1982 Dynamics of phase transition in the new inflationary universe scenario and generation of perturbations. {\it Phys. Lett. B} {\bf 117}, 175.

\item Steidel, C. 1998 in this volume.

\item Turner, M.S. 1993 Recovering the inflationary potential. {\it Phys. Rev. D} {\bf 48}, 5539.

\item Turner, M.S. 1993 Dark matter: Theoretical perspectives. {\it Proc. Natl. Acad. Sci.} {\bf 90}, 4827.

\item Turner, M.S. 1997a. In {\it Generation of cosmological large-scale structure} (eds. D.N. Schramm and P. Galeotti),  p.153. Dordrecht:Kluwer.

\item Turner, M.S. 1997b The case for $\Lambda$ CDM. In {\it Critical dialogues in cosmology} (ed. N. Turok), p. 555. Singapore:World Scientific. 

\item Turner, M.S. 1997c Detectability of inflation-produced gravitational waves. {\it Phys. Rev. D} {\bf 55}, R435.

\item Turner, M.S., Steigman, G. \& Krauss, L. 1984 The `flatness of the universe': Reconciling theoretical prejudice with observational data.  {\it Phys. Rev. Lett.} {\bf 52}, 2090.

\item Turner, M.S. \& White, M. 1996 {\it Phys. Rev. D} {\bf 53}, 6822.

\item Turner, M.S. \& White, M. 1997 CDM models with a smooth component. {\it Phys. Rev. D} {\bf 56}, R4439.

\item Turok, N. 1998 in this volume.

\item Vilenkin, A. 1984 String dominated universe. {\it Phys. Rev. Lett.} {\bf 53}, 1016-1018.

\item White, S.D.M.,  Frenk, C., \& Davis, M. 1983 Clustering in a neutrino-dominated universe. {\it Astrophys. J.}
{\bf 274}, L1-L6.

\item Willick, J. A., Strauss, M. A., Dekel, A., \& Kolatt, T., 1997
Maximum likelihood comparisons of Fisher-Tully
and redshift data: constraints on $\Omega$ and biasing.
{\it Astrophys. J.} {\bf 486}, 629.

\end {thebibliography}
\end {document}